# SIM-CE: An Advanced Simulink Platform for Studying the Brain of Caenorhabditis elegans

**Ramin M. Hasani** [1] **Victoria Beneder** [2] **Magdalena Fuchs** [1] **David Lung** [1] **Radu Grosu** [1]

## Abstract

We introduce SIM-CE, an advanced, user-friendly modeling and simulation environment in Simulink for performing multi-scale behavioral analysis of the nervous system of Caenorhabditis elegans (*C. elegans*). SIM-CE contains an implementation of the mathematical models of *C. elegans*'s neurons and synapses, in Simulink, which can be easily extended and particularized by the user. The Simulink model is able to capture both complex dynamics of ion channels and additional biophysical detail such as intracellular calcium concentration. We demonstrate the performance of SIM-CE by carrying out neuronal, synaptic and neural-circuit-level behavioral simulations. Such environment enables the user to capture unknown properties of the neural circuits, test hypotheses and determine the origin of many behavioral plasticities exhibited by the worm.

## 1. Introduction

*C. elegans* is most likely the world's best-understood animal (Ardiel & Rankin, 2010). However, the fundamental principles underlying its behavior are yet to be understood. Its relatively simple nervous system is constructed from 302 neurons, hardwired by means of about 8000 chemical and electrical synapses (Varshney et al., 2011). Despite its simplicity, it has shown remarkable complex behavioral features which make it an attractive model system. Multi-scale behavioral analyses have been conducted on the worms nervous system (Kato et al., 2015; Roberts et al., 2016), and global attempts on modeling its emergent behavior have been commenced (Szigeti et al., 2014). In this regard, there is a high demand for a comprehensive neuron-by-neuron model platform which incorporates many electrophysiological properties of neurons while avoiding large parameter space (Roberts et al., 2016; Hasani et al., 2016).

In the present study, we construct a modular simulation platform, SIM-CE, for investigating the fundamental principles underlying physiological processes within the neural circuits of the brain of *C. elegans*, in Simulink. There are three key components in the process of modeling the brain dynamics; 1) The choice of a suitable model for imitating the desired behavior and analysis. 2) Deciding the level of abstraction and biophysical details to be considered. 3) selecting a proper model for the connectivity of the network. Accordingly, we design Simulink models of single-compartmental neuron models, as previously developed in (Kuramochi & Iwasaki, 2010), and synapses, and employ them in the implementation of neural circuits. In particular, we exemplify the performance of our modeling platform by creating the tap-withdrawal (TW) neural circuit, a circuit which mediates a reflexive response to mechanical stimulation of touch sensory neurons (Wicks & Rankin, 1997).

We show that SIM-CE can capture behavioral features of the neural circuits in multi-scale details from biophysics of neurons and synapses. Moreover, SIM-CE can presumably test hypotheses and provide valid predictions. For instance one can define the structure of the wiring diagram, reveal the working principles of numerous unknown neural circuits such as the central pattern generator (CPG), determine the synaptic polarities and find the origin of many associative and non-associative learning modalities (Hasani et al., 2017).

We structure the paper as follows. In Section 2, we step-by-step explain the design procedures of SIM-CE platform. We recapitulate the mathematical modeling methods and show how to add various ion channel dynamics to the neuron model, and complete the Simulink design in Section 2.1. We also illustrate how to design the connectivity among the neurons in Section 2.2. We then, in Section 2.3, demonstrate the capability of the designed platform to construct neural circuits and correspondingly simulate their dynamics. We provide a useful discussion on the attributes of SIM-CE platform in Section 3 and conclude our work in Section 4.

[1]Vienna University of Technology, Austria [2]University of Natural Resources and Life Sciences, Vienna, Austria. Correspondence to: Ramin M. Hasani <ramin.hasani@tuwien.ac.at>.

## 2. Methods

In this section we introduce the methodologies utilized for implementation of Simulink models of the neurons and synapses. We revisit the single-compartmental neuron modeling techniques initially introduced by Hodgkin & Huxley (1952) and Kuramochi & Iwasaki (2010). The model comprises several ordinary differential equations (ODE) describing dynamics of the cell. ODEs are implemented by using the integrator block in Simulink. For instance, the equation $\frac{dV}{dt} = -60 - V$, is constructed as shown in Figure 1. The input to the integrator block is the derivative term $dv/dt$. The output is $V(t)$, which is multiplied by $-1$, summed up with the constant and is fed back to the integrator's input. The initial condition of the variable $V$ is defined inside the integrator block. The equation is then solved by using the ode45 solver of MATLAB which realizes the Dolman-Prince (Dormand & Prince, 1980), a numerical method for solving ODEs.

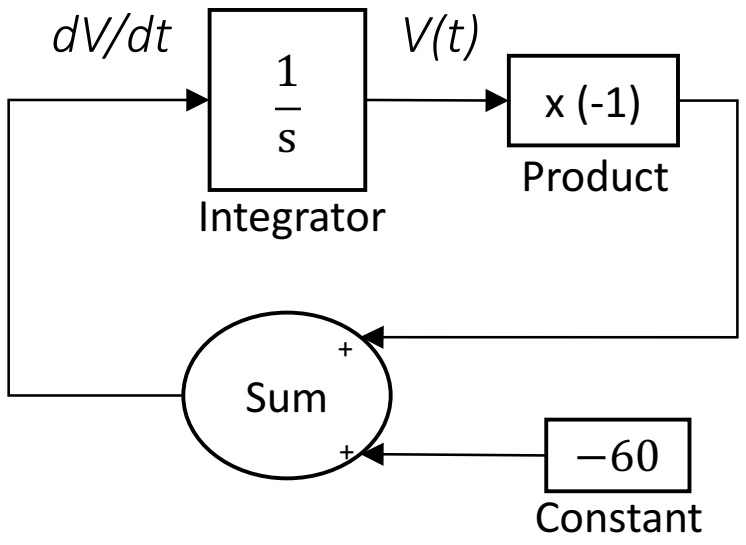

*Figure 1.* ODE implementation in Simulink

### 2.1. Neuron Model

The overall current passing through the membrane of a neuron is maintained by the sum of inward and outward currents passing through it. Such currents are generated as a result of propagation of ions across the cell membrane. Figure 2 shows the membrane of a *C. elegans* neuron where various channels such as voltage-gated potassium channels, calcium-gated potassium channels, voltage-gated calcium channels and leakage channels together with a calcium pump establish the membrane potential of the neuron.

Accordingly, the dynamics of the membrane potential can be written in the form of the following differential equation:

$$C_m \frac{dV}{dt} = -(I_{Ca} + I_K + I_{sk} + I_{Leak}) + \Sigma(I_{Syn} + I_{gap} + I_{stimuli}), \quad (1)$$

where $C_m$ corresponds to the membrane capacitance and $V$ is the membrane potential. $I_{Ca}$, $I_K$, $I_{sk}$ and $I_{Leak}$ represent the calcium current, potassium current, intercellular calcium-gated potassium channel current and leakage current respectively (Kuramochi & Iwasaki, 2010). $I_{stimuli}$, $I_{syn}$ and $I_{gap}$ are currents which are injected into the neuron from a stimulus to a sensory neuron, a chemical synapse and an electrical synapse (gap junction) respectively.

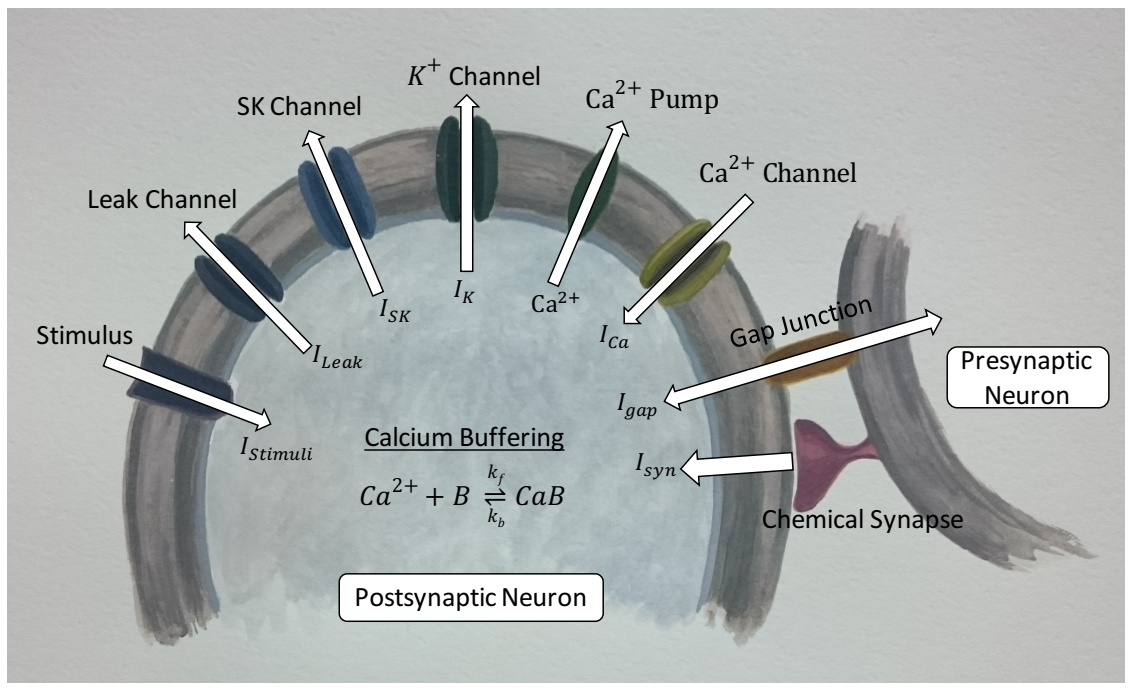

*Figure 2.* Representation of a single neuron of *C. elegans* (Painting by: suppressed for anonymity, reproduced from (Kuramochi & Iwasaki, 2010)).

For modeling the ion conductance channels we use the classic Hodgkin-Huxley formalism (Hodgkin & Huxley, 1952) where the conductance of a channel, $G_{ion}$, is defined by the product of the maximum conductance, the probability of activation function, $m$, and inactivation function, $h$, of the channel as follows:

$$G_{ion} = G_{max} m^p h^q. \quad (2)$$

For the calcium and potassium channels we only consider the activation variables as proposed by Hodgkin & Huxley (1952) and Engel et al. (1999). We define $\alpha$ and $\beta$, the variables of the gate-rate-functions $m$ for each channel by analyzing the steady-state of the function, $m_\infty$, and the time constant, $\tau_m$, of the gate relaxation (Hodgkin & Huxley, 1952).

$$m_\infty = \alpha_m / (\alpha_m + \beta_m) \quad (3a)$$

$$\tau_m = 1/(\alpha_m + \beta_m) \quad (3b)$$

Once we determined the gate-rate-functions for each channel, we describe the straightforward way of implementing each channel and consequently the neuron itself, as below.

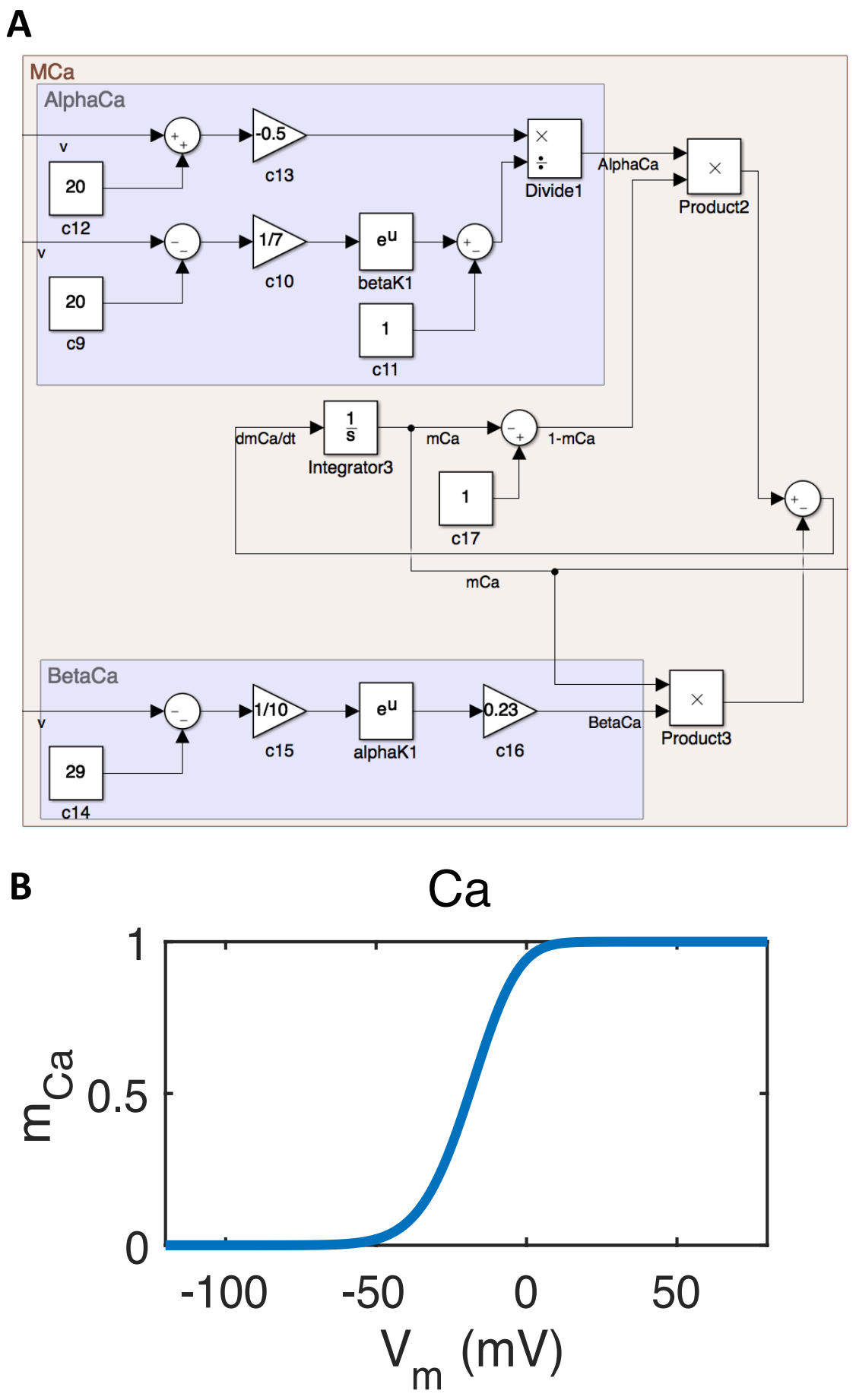

*Figure 3.* Calcium gate-rate-function. A) Simulink implementation of calcium gate-rate-function. B) Activation function of the calcium conductance.

### 2.1.1. Calcium Current

Calcium ions are key elements in signalling in the nervous systems. They are responsible for several functions such as synaptic vesicle release, activation of ionic channels and excitatory currents across the membrane of neurons (Hobert, 2013). This is notable since *C. elegans* lacks voltage-gated sodium channels and hence it does not generate sodium-based action potentials (Goodman et al., 1998). The current passing through the voltage-sensitive calcium channel can be derived from the Goldman-Hodgkin-Katz equation (Engel et al., 1999) as follows:

$$I_{Ca} = m_{Ca}^2 P_{Ca} \frac{n_{Ca} F V}{RT} \frac{[Ca^{2+}]_{in} - [Ca^{2+}]_{out}\, e^{\frac{-2FV}{RT}}}{1 - e^{\frac{-2FV}{RT}}}, \tag{4}$$

where $P_{Ca}$, $n_{Ca}$, $F$, $R$, $T$ and $[Ca^{2+}]_{out}$ are calcium permeability, calcium charge valance, Faraday constant, gas constant, temperature (in Kelvin) and extracellular calcium concentration (considered constant), respectively (Kuramochi & Iwasaki, 2010). Later on, we employ three different mechanisms in order to set the value of the concentration of calcium inside the neuron, $[Ca^{2+}]_{in}$.

As we discussed before, the dynamics of the voltage dependent gating variable, $m_{Ca}$, can be derived from equations 3a and 3b:

$$\frac{dm_{Ca}}{dt} = \alpha_{Ca}(V)(1 - m_{Ca}) - \beta_{Ca}(V) m_{Ca}, \tag{5a}$$

where

$$\alpha_{Ca}(V) = \frac{-0.5(V + 20)}{e^{\frac{-(V+20)}{7}} - 1} f \tag{5b}$$

and

$$\beta_{Ca}(V) = 0.23 e^{\frac{-(v+29)}{10}}. \tag{5c}$$

We construct the behavior of $m_{Ca}$ based on equation 5a in Simulink as shown in Figure 3A. The variation of the activation function of the calcium conductance as a function of the membrane potential is plotted in Figure 3B.

### 2.1.2. Potassium Current

In almost all animal cells, potassium channels are responsible for setting the resting potential and bringing a neuron from an excited state back to its equilibrium state (Katz, 1966). The concentration of potassium ions inside the cell is higher than outside, therefore, potassium tends to flow outward. The current generated by the outflow of potassium ions, can be abstracted as follows:

$$I_K = G_K n_K^4 (V - E_K), \tag{6}$$

where $G_K$ and $E_K$ are the maximum potassium channel conductance and potassium equilibrium potential, respectively. The kinetics of the potassium gating variable, $n_K$, is given by Equation 7a, which is directly derived from Equations 3a and 3b (Hodgkin & Huxley, 1952):

$$\frac{dn_K}{dt} = \alpha_K(V)(1 - n_K) - \beta_K(V) n_K. \tag{7a}$$

where $\alpha_K(V)$, the gating variable which is changing with the presynaptic membrane potential, is calculated as:

$$\alpha_K(V) = 0.6 e^{\frac{v+40}{20}} \tag{7b}$$

and $\beta_K(V)$ is computed as follows:

$$\beta_K(V) = \frac{0.01(V + 12)}{e^{\frac{V+12}{23}} - 1}. \tag{7c}$$

We implement the described potassium-conductance activation function in Simulink as illustrated in Figure 4A and show its dynamics as a function of the membrane potential in Figure 4B. Note that the activation of the potassium channel is set to be slightly slower than that of the calcium channels.

### 2.1.3. SMALL-CONDUCTANCE POTASSIUM CHANNEL'S CURRENT

As regulators, these channels significantly influence the excitability of the neurons (Salkoff et al., 2005). Small-conductance calcium-activated potassium channels are also included as it is shown in Equation 8.

$$I_{sK} = G_{sK} m_{sK} (V - E_{sK}), \tag{8}$$

where $G_{sK}$ and $E_{sK}$ show the maximum channel conductance and the corresponding Nernst potential, respectively. Activation of these channels is purely controlled by the intracellular calcium concentration in the form of the following equation (Engel et al., 1999):

$$\frac{dm_{sK}}{dt} = \frac{[Ca^{2+}]_{in}}{\psi_{sK}}(1 - m_{sK}) - \frac{K_{sK}}{\psi_{sK}} m_{sK}, \tag{9}$$

where $\psi_{sK}$ as the time constant rate and $K_{sK}$ as the potassium concentration are used in order to calculate the gated variable rate functions $\alpha$ and $\beta$, in a classic Hudgkin-Huxley scheme (Kuramochi & Iwasaki, 2010). We design the Simulink model of the small-conductance potassium channel as it is depicted in Figure 5A and show its dynamics as a function of the inner calcium concentration in Figure 5.

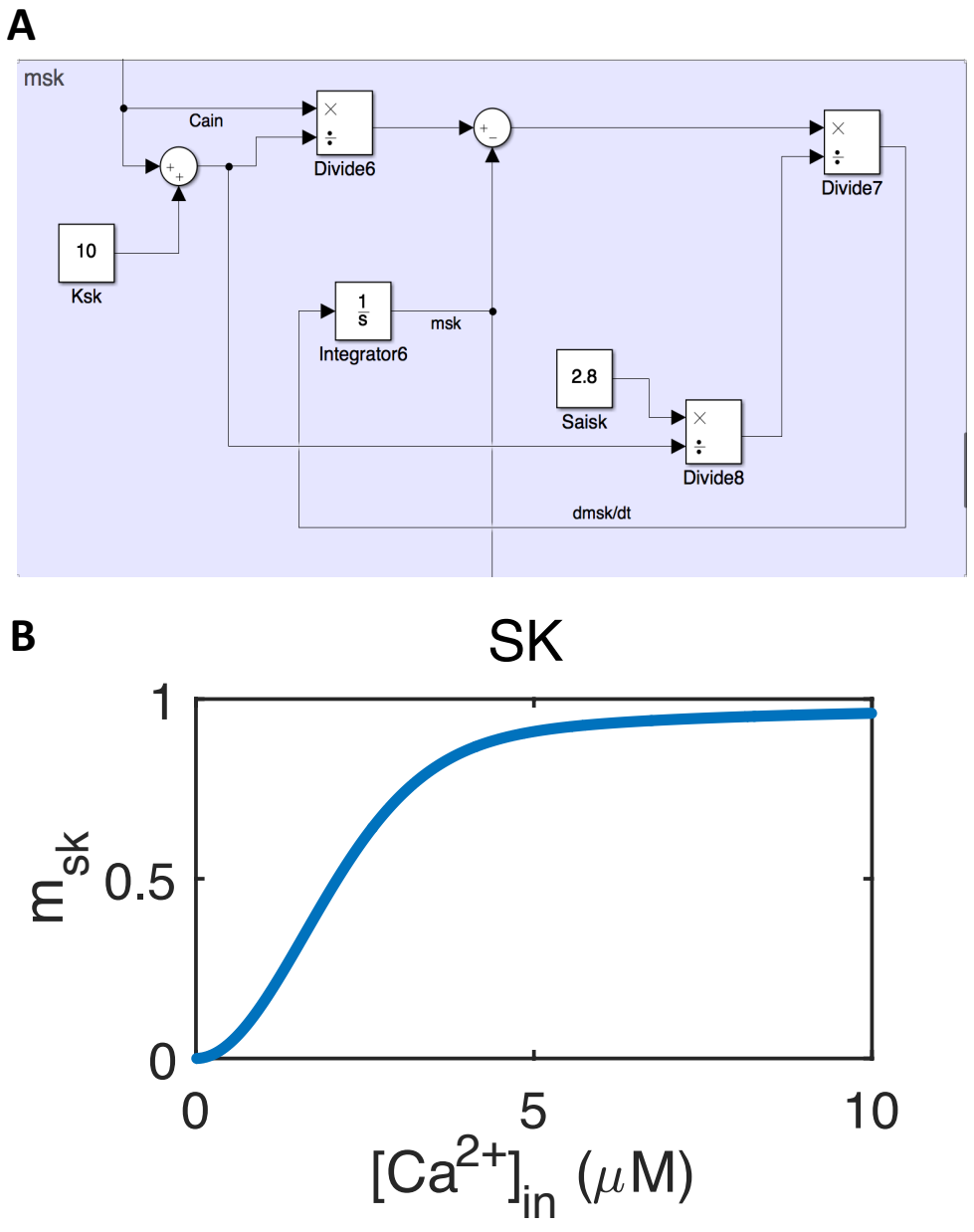

*Figure 5.* Small-conductance potassium channel (SK) gate-rate-function. A) Simulink model of the SK B) Activation function of the potassium conductance as a function of the intracellular calcium concentration.

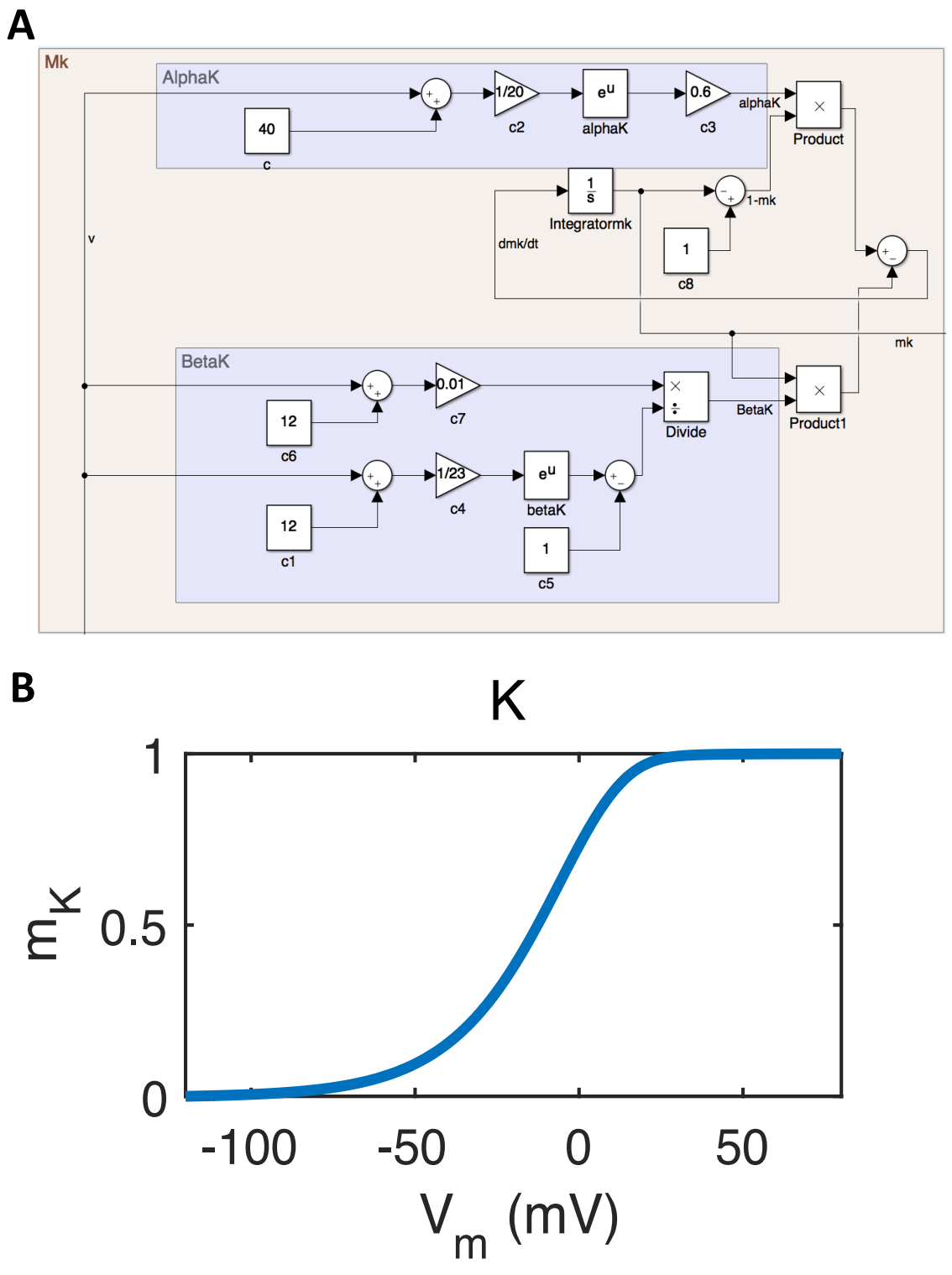

*Figure 4.* Potassium channel gate-rate-function. A) Simulink implementation B) Activation function of the potassium conductance.

### 2.1.4. LEAKAGE CURRENT

Leakage channels are the mediators of the cells where ions can freely move along them. For a neuron we model the leakage current passing through a leakage channel with a constant conductance, $G_{Leak}$, and an average Nernst potential of $E_{Leak}$, as follows:

$$I_{Leak} = G_{Leak}(V - E_{Leak}). \tag{10}$$

Figure 6 shows the Simulink model of the leakage channel.

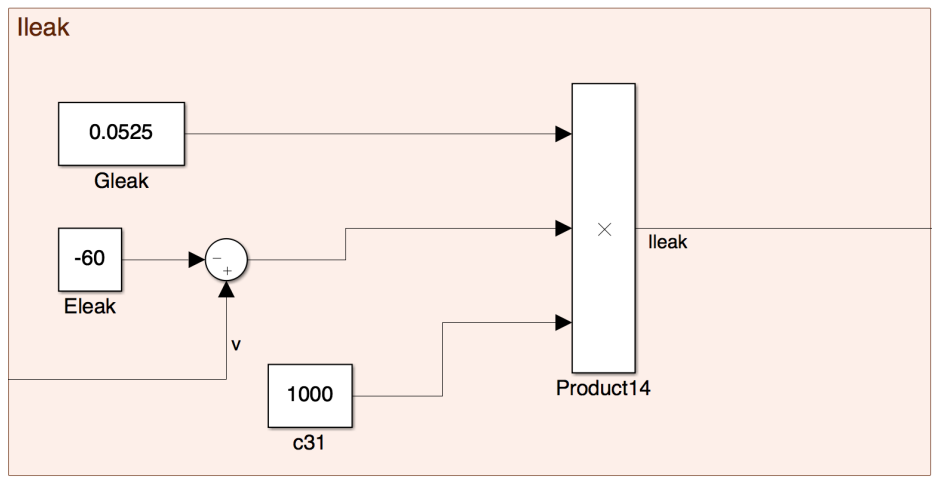

*Figure 6.* Simulink model of the leakage channel

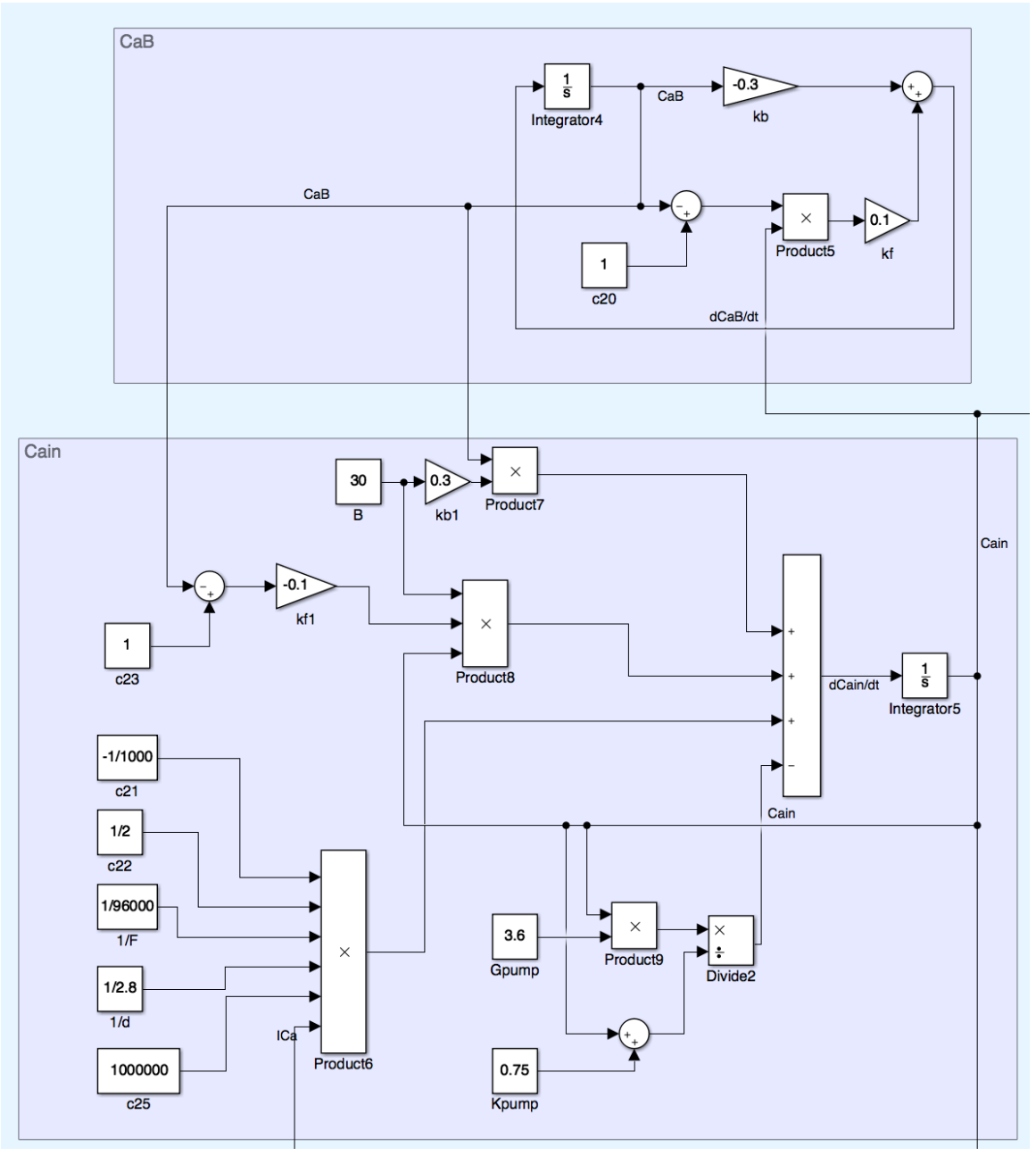

*Figure 7.* Simulink model of the inner calcium concentration

### 2.1.5. Intracellular Calcium Concentration Dynamics

During the last years the use of calcium imaging resulted in significant advancements in the understanding of the physiology of neurons and neural circuits of *C. elegans* (Kato et al., 2015). The dynamics of the intracellular calcium concentration for a neuron represents the dynamics of the whole cell. There are several mechanisms involved in determining the inner calcium level; Inflow of the calcium current through voltage-gated calcium channels is the main source of an increase in its level. Calcium ions inside the cell bind to specific types of proteins called binding proteins. Such chemical process varies the amount of freely available intracellular calcium ions (Kuramochi & Iwasaki, 2010). Moreover, a calcium pump gets activated when the amount of intracellular calcium exceeds a certain threshold and transports calcium out of the cell. In *C. elegans*, calcium currents play the excitatory role that sodium currents have in mammalian cells. Dynamics of the inner calcium concentration based on the described mechanisms is derived in the Equation 11a where $d$ is the calcium diffusion length, $k_b$, $k_f$ are the coefficients of the bidirectional chemical process of the binding proteins $[B]$, with calcium ions which create the molecule, $[CaB]$ and $G_{pump}$ is the conductance of the pump (Kuramochi & Iwasaki, 2010). Equation 11b quantitatively represents the dynamics of the chemical process of the calcium ions and binding proteins (Kuramochi & Iwasaki, 2010).

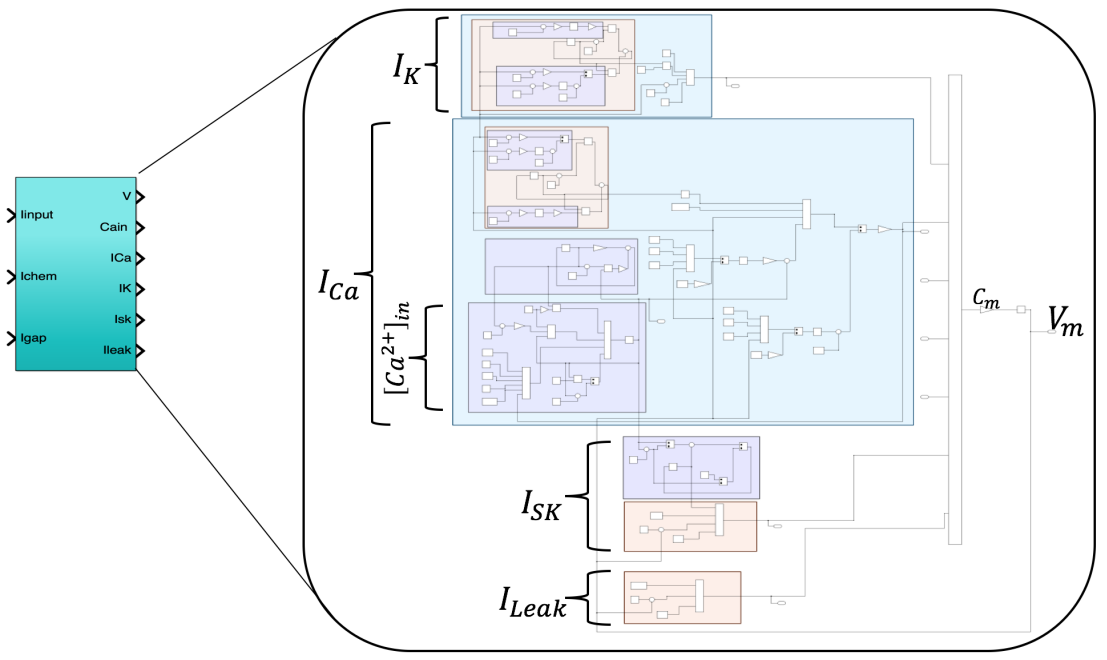

*Figure 8.* Simulink model of a *C. elegans* neuron

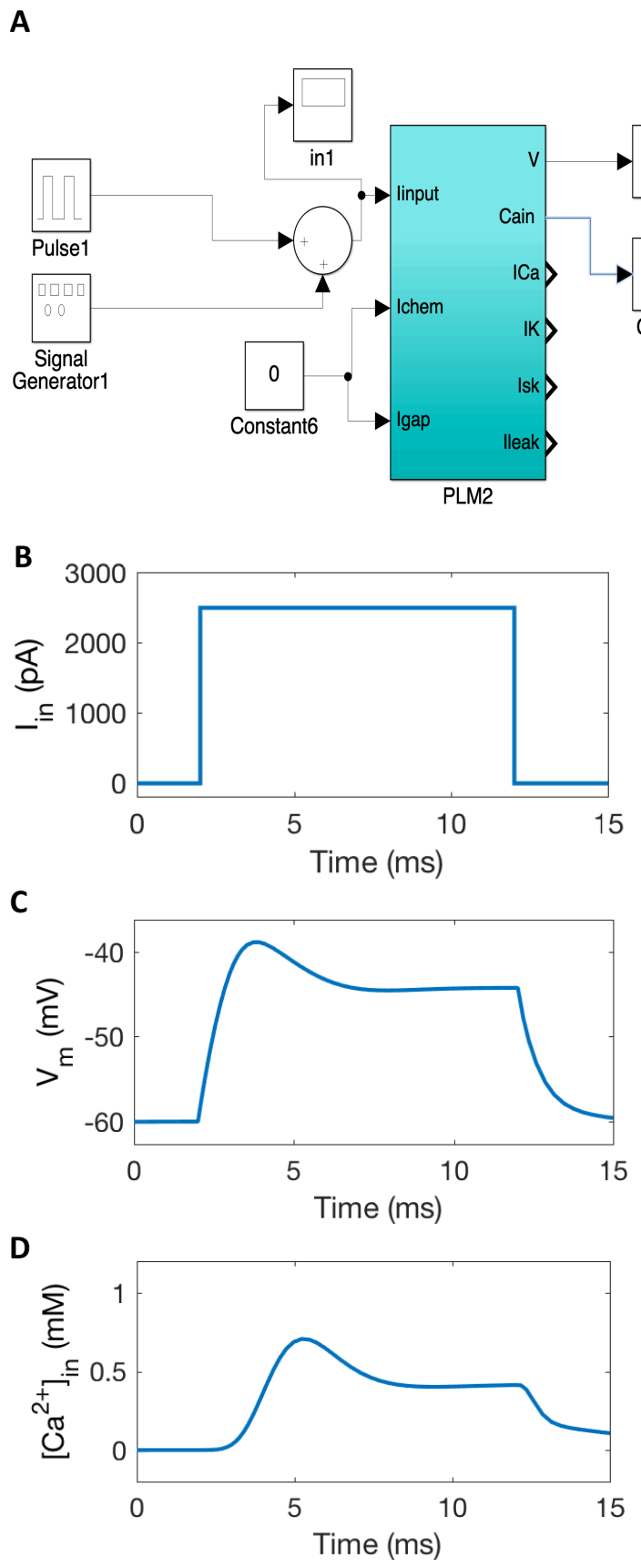

*Figure 9.* Simulation of the deterministic response of the neuron to an input current stimulus.

$$\frac{d[Ca^{2+}]_{in}}{dt} = -\frac{1}{2Fd}I_{Ca} + k_b[B][CaB] - k_f[Ca^{2+}]_{in}[B](1-[CaB]) - \frac{G_{pump}[Ca^{2+}]_{in}}{[Ca^{2+}]_{in} + K_{pump}} \tag{11a}$$

$$\frac{d[CaB]}{dt} = -k_b[CaB] + k_f[Ca^{2+}]_{in}(1-[CaB]) \tag{11b}$$

Intracellular calcium concentration kinetics is designed as a part of the neuron model and shown in Figure 7.

All the building blocks of the neuron model shown in Figure 2 are designed and put together, forming a single neuron block. Figure 8 represents the schematic of such detailed neuron model where all the parameters are conveniently adjustable. The model enables the user to input external stimuli from the environment or from a presynaptic synapse in real time while monitoring the physiological behavior of the membrane potential, intracellular calcium concentration and the ion channel currents. Note that there are several types of ion channels determined in the neurons of *C. elegans* (Salkoff et al., 2005). One can easily implement models of any arbitrarily chosen ion channel and include it in the neuron dynamics. Table 1 summarizes the parameters of the neuron and synapse model together with the range of their values.

*Table 1.* Parameter-space of the model of neuron and synapse

| PARAMETER | VALUE-RANGE | PARAMETER | VALUE-RANGE |
|---|---|---|---|
| $C_m$ | $15 - 100\ \frac{\mu F}{cm^2}$ | $G_{Leak}$ | $0.04 - 0.1\ \frac{mS}{cm^2}$ |
| $P_{Ca}$ | $1 - 3\ \frac{nA}{\mu M.mV.cm^2}$ | $E_{Leak}$ | $-74mV$ |
| $n_{Ca}$ | $2$ | $d$ | $1 - 4\ \mu m$ |
| $F$ | $9.6 \times 10^4\ \frac{J}{M.V}$ | $k_b$ | $0.3\ \frac{1}{ms}$ |
| $R$ | $8.3\ \frac{J}{M.^{\circ}K}$ | $[B]$ | $15 - 30\ \mu M$ |
| $T$ | $293.1\ ^{\circ}K$ | $k_f$ | $0.1\ \frac{1}{ms.\mu M}$ |
| $[Ca^{2+}]_{out}$ | $2 - 2.4\ mM$ | $K_{pump}$ | $0.3 - 1\ \mu M$ |
| $G_K$ | $1 - 15\ mS/cm^2$ | $G_{pump}$ | $1 - 10\ \frac{\mu M}{ms}$ |
| $E_K$ | $-95\ to\ -60\ mV$ | $E_{syn}$ | $-90 - 0\ mV$ |
| $G_{sK}$ | $0.2 - 0.3\ mS/cm^2$ | $G_{syn}$ | $0.1 - 1\ \frac{mS}{cm^2}$ |
| $E_{sK}$ | $-95\ to\ -60\ mV$ | $V_{shift}$ | $-10\ to\ -40\ mV$ |
| $\psi_{sK}$ | $2.5 - 3\ \mu M.ms$ | $V_{range}$ | $3 - 6\ mV$ |
| $K_{sK}$ | $0.1 - 0.9\ \mu M$ | $G_{gap}$ | $0.1 - 1\frac{mS}{cm^2}$ |

We now simulate a single neuron and observe its response to an external stimulus. Figure 9 demonstrates the deterministic response of the designed neuron to an external input stimulus shown in Figure 9B. The membrane potential response together with the intracellular calcium concentration is plotted in Figure 9C and 9D, respectively. Calcium dynamics follows the excitability of the membrane potential demonstrating how calcium imaging can be representative of the dynamics of the whole cell.

In addition to reproducing the deterministic behavior of the neuron, we include stochastic properties to the system. We apply a Gaussian white noise on every ion channel conductance and also consider the intrinsic random activation of the neuron by stimulating the neuron with a random current-pulse generator. The results of a transient simulation are shown in Figure 10 where the membrane potential together with the calcium concentration of a single neuron is plotted in Figure 10B and 10C, respectively. The calcium dynamics is similar to that of biological neurons captured in calcium imaging experiments with a reasonable degree of accuracy (Kato et al., 2015). In this way, the model additionally allows the user to study the intrinsic stochastic behavior of the cells and neural circuits (Roberts et al., 2016).

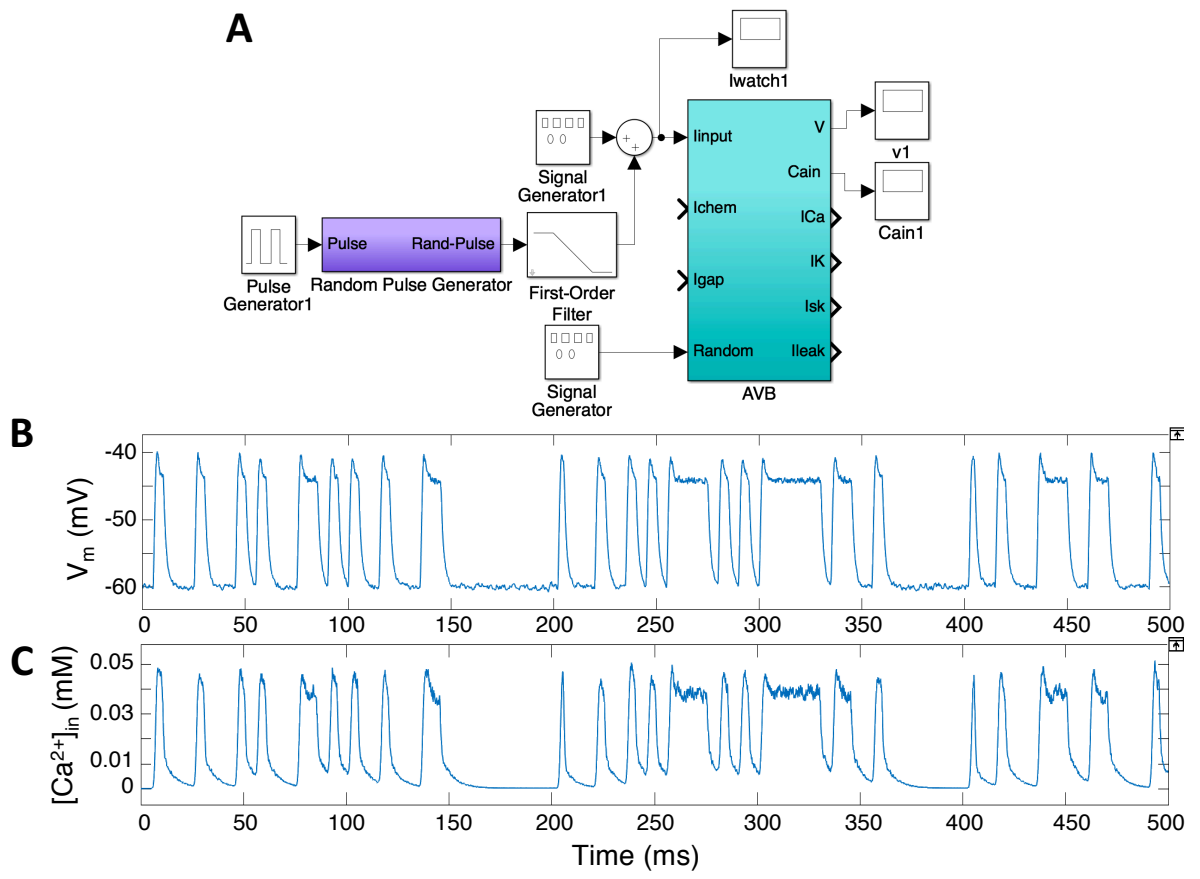

*Figure 10.* Simulation of the stochastic response of the neuron to an intrinsic random current-pulse generator. A) The simulation setup. B) Membrane potential C) Calcium concentration.

## 2.2. Wiring

Synapses are the points of communication of neurons. Information transfer happens as a result of neurotransmitters release at a chemical synaptic port of a neuron. There are two major types of synapses, chemical synapses and electrical synapses (gap junctions). Chemical synapses are the elements at which the signaling from a presynaptic neuron to a postsynaptic cell occurs through secretion of neurotransmitters. Depending on the type of neurotransmitters available at the presynaptic neuron, a chemical synapse can be excitatory or inhibitory for the postsynaptic neuron. Dynamics of the chemical synapses depends on several factors, such as the concentration of the available neurotransmitters, the state of the activation/inactivation of the previous neuron, the probability of the binding of neurotransmitters to the postsynaptic receptors and the amount of the receptors available at the postsynaptic cell (Schutter, 2009).

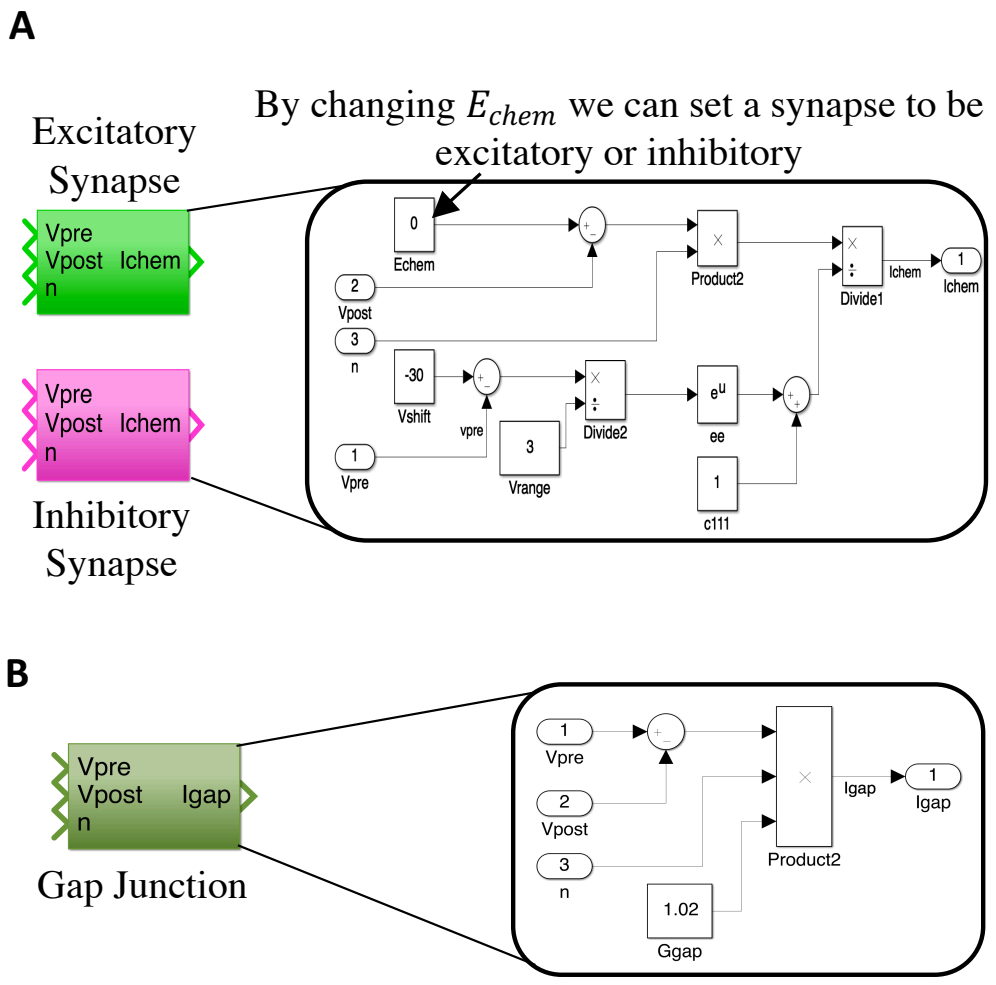

*Figure 11.* Simulink model of synapses. A) Chemical excitatory and inhibitory synapse model B) Model of the gap junction.

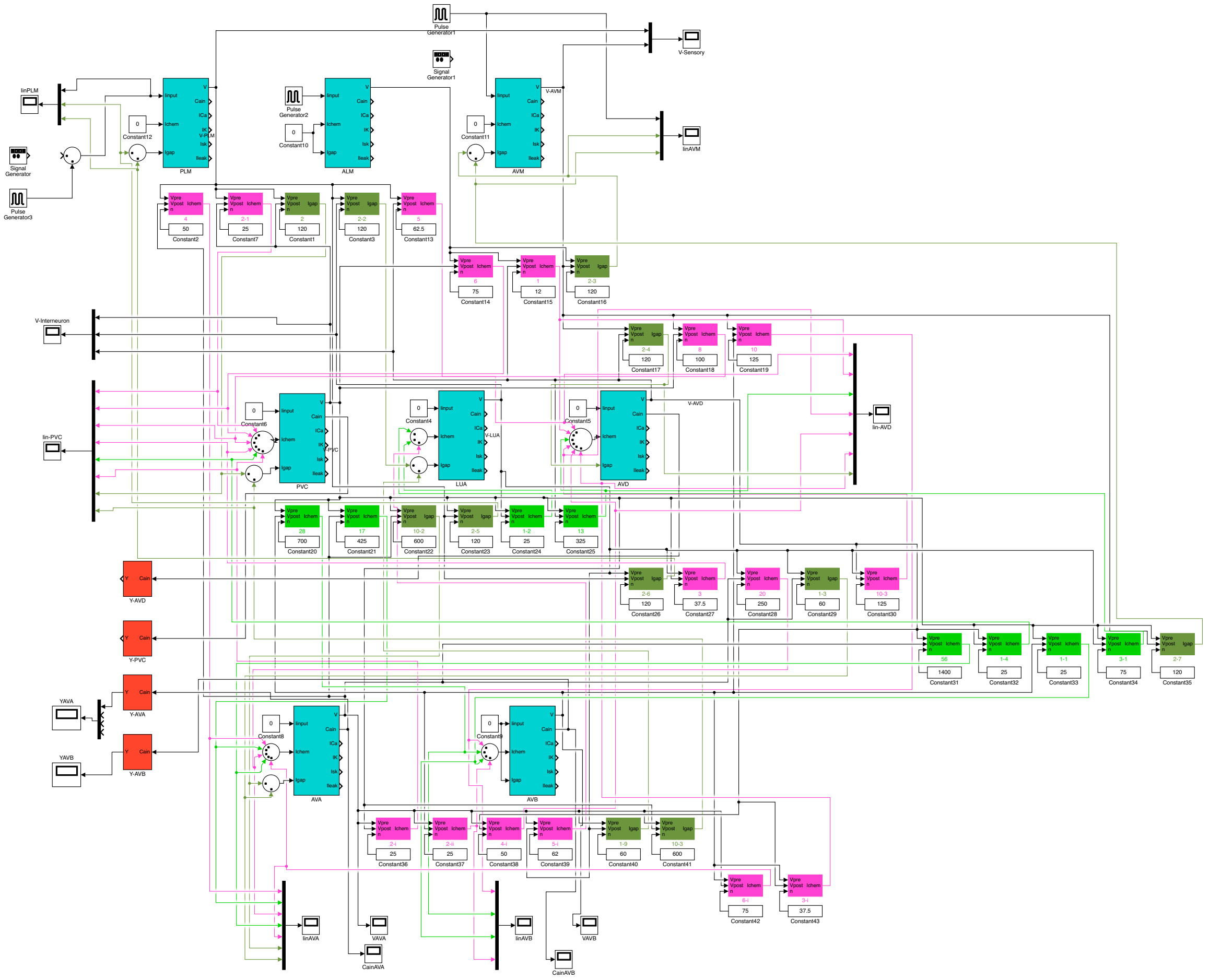

*Figure 12.* Tap-withdrawal neural circuit implemented in the Simulink platform.

A static mathematical representation of such synapses can be formulated as follows:

$$I_{Syn} = n_{ij} \frac{G_{Syn}}{1 + e^{\frac{-(V_{pre} - V_{shift})}{V_{range}}}} (E_{Syn} - V_{post}), \qquad (12)$$

where, $n_{ij}$ is the number of synaptic connections from neuron $i$ to neuron $j$, $G_{syn}$ is the maximum conductance of the synapse, $V_{shift}$ and $V_{range}$ stand for the equilibrium potential and the steepness of the synaptic activation, respectively (Koch & Segev, 1998). $E_{Syn}$, represents reversal potential of the synapse where its variations results in excitatory or inhibitory behavior of the synapse. In our model it is set to zero for an excitatory synapse and to $-90mV$ for an inhibitory synapse. Figure 11A represents the designed Simulink model of chemical synapses.

Electrical synapses (gap junctions) can be modeled simply by employing Ohm's law as follows:

$$I_{gap} = n_{gap} G_{gap} (V_{pre} - V_{post}), \qquad (13)$$

where $n_{gap}$ and $G_{gap}$, represent the number gap junctions between two neurons and the gap junction conductance, respectively. Figure 11B depicts the implementation of the gap junction in Simulink.

## 2.3. Neural Circuits Implementation

In this section we illustrate how our proposed Simulink platform can enable users to construct arbitrarily chosen neural circuits from the nervous system of the worm and simulate its behavior while fully observing the dynamics of the circuit in neuronal and synaptic level.

An example of a well-understood neural circuit within the nervous system of *C. elegans*, is the tap-withdrawal (TW) circuit. The circuit modulates a reflexive response to a mechanical stimulus subjecting the petri dish in which the worm crawls. TW circuit comprises 8 neurons optimally hardwired through almost 360 synapses. A mechanical stimulus excites specific sensory neurons (PLM, AVM or ALM) and results in the activation of the corresponding command neurons (AVA or AVB) through a group of interneurons (PVC, LUA, AVD). AVA activates the motor neuron responsible for reversal movement while AVB command neuron initiates forward locomotion. Depending on the strength of the stimulation on the sensory neurons, AVA or AVB gets activated and results in a reflexive motion (Wicks & Rankin, 1997).

Here, we construct the model of the neural circuit in our platform. Figure 12, represents the designed TW circuit. The architecture of the circuit is taken from the initial *C. elegans* connectome data (White et al., 1986). The polarity of the synaptic connections is set based on the evaluations provided in (Wicks & Rankin, 1997).

We then perform a simulated experiment in which we stimulate the AVM sensory neuron and show how the command neurons behave accordingly. Figure 13 shows the behavior of the command neurons AVA and AVB to the described stimulation. Figure 13B and 13C, show the membrane potential of the command neurons and Figure 13D and 13E depict their intracellular calcium concentrations. We observe that the response of the model correctly imitates the expected behavior where an anterior stimulation of the worm results in the activation of the reversal command neuron and therefore initiation of a backward reflexive movement. Note that the forward-movement command neuron, AVB, is suppressed due to the dynamics of the circuit.

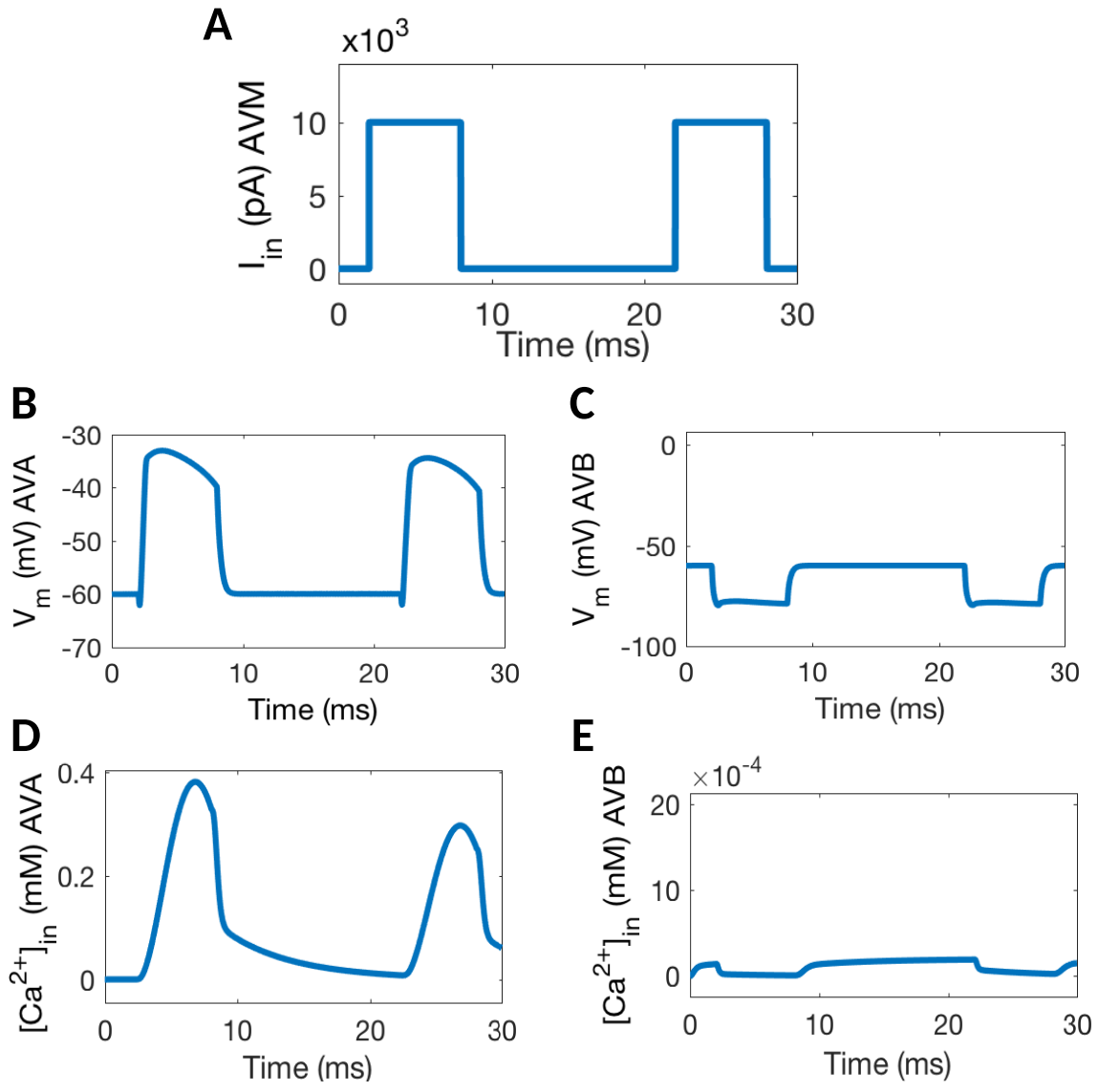

*Figure 13.* Responses of the tap-withdrawal neural circuit A) Inputs to the sensory neurons B) Output of the AVA command neuron C) Output of the AVB comannad neuron D) Intracellular calcium concentration of the AVA neuron E) Intracellular calcium concentration of the AVB neuron

The top-level block diagram representation in SIM-CE considerably simplifies the investigation of basic behavioral primitives within the C.elegans neural circuits. Moreover, SIM-CE is a suitable base for performing large-scale block level simulations. This enables our platform to be flexible regarding simulation of large-scale networks (ideally the entire nervous system). However, simulation of such networks would essentially require much hardware computational power.

## 3. Discussion

Over the past years a considerable number of behavioral analyses on the nervous system of *C. elegans* have been conducted. However, the origin of many behavioral plasticities have not yet been fully determined. SIM-CE platform, provides researchers with the opportunity to explore beyond the observable dynamics in the brain of the worm where multi-scale details of the neural circuits can be constructed and decoded, precisely in a user-friendly fashion. SIM-CE permits us to study attractive behavioral features of *C. elegans* such as gene modification effects, non-associative and associative learning in various levels by tuning parameters (**?**). Moreover, one can perform numerous simulated experiments within the platform and provide multi-scale predictions on the physiology of the neural circuits.

## 4. Conclusions

We introduced SIM-CE, a Simulink-based modeling and simulation platform for investigating the nervous system of *C. elegans*. The platform offers detailed mathematical models of the neurons and synapses and enables the user to conveniently perform simulated experiments in neuronal, synaptic and neural circuit level. We comprehensively explained the design-flow of single neurons and synapses. We then exemplified the performance of the platform by performing simulated experiments of single neurons in deterministic and stochastic modes, and a neural circuit.

For future work, we aim to explore the dynamics of more neural circuits in multi-scale physiological conditions, provide valuable predictions about the unknown properties of such circuits and ultimately include the dynamics of the entire nervous system of the worm.